\documentstyle[12pt,epsf]{article}

\newcommand{\sigmac}{\bar{\psi}\psi}
\newcommand{\pic}{\bar{\psi}i\gamma_5\vec{\tau}\psi}
\newcommand{\bvec}[1]{\mbox{\boldmath $#1$}}

\begin{document}

\title{Color superconductivity at finite density\\
and temperature with flavor asymmetry}

\author{O. Kiriyama, S. Yasui, and H. Toki\\
{\small {\it Research Center for Nuclear Physics, Osaka University,
Ibaraki 567--0047, Japan}}}
\date{RCNP-Th01012}
\maketitle

\begin{abstract}
Color superconductivity of QCD at finite density, temperature and 
flavor asymmetry is studied within an approximation which the interaction 
is modeled upon four--fermion interactions. 
We calculate the thermodynamic potential 
in the so-called NJL-type model and study the phase structure 
at finite density, temperature and flavor asymmetry. 
We find that a mixed phase appears at sufficiently large 
flavor asymmetry and low temperature. A tricritical point 
in the $T$-$\mu_B$-$\mu_I$ plane is also found.
\end{abstract}

\section{INTRODUCTION}

Quantum Chromodynamics (QCD) describes the interaction of quarks 
and gluons. Since QCD is asymptotically free, the coupling constant 
becomes small when either temperature or the chemical potential, namely 
the the Fermi momentum, is large. 
By Cooper theorem, Fermi surface has instability when an arbitrarily 
small attractive interaction between quarks presents. 
For quark--quark scattering, where the quarks are in 
the antitriplet $\bvec{\bar{3}}$, one gluon exchange interaction 
is attractive and in the $\bvec{6}$, it is repulsive. 
Thus we expect that, in the case of two quark flavors, 
the quark--quark condensate in $\bvec{\bar{3}}$ 
channel is formed and the color superconductivity, spontaneous 
breakdown of the color gauge symmetry $SU(3)_C \to SU(2)_C$, 
occurs \cite{csc}. 
This phenomenon can be understood in the sense that the 
effective dimensional reduction $(3+1) \to (1+1)$ occurs 
with the large chemical potential 
as well as the case of $\mbox{QED}_4$ in a constant magnetic field \cite{qed}.

There has been many studies of the color superconductivity 
in cold and dense QCD using 
the so-called Nambu--Jona-Lasinio (NJL) type 
models \cite{rajagopal,klevansky}, 
the instanton liquid model \cite{shuryak} and 
the Schwinger-Dyson approach to QCD in weak coupling limit \cite{oge}, 
in other words, in $\mu\to\infty$ limit. 
The first two models have shown that the superconducting gap 
could be large as 100 MeV. 
On the other hand, due to the exchange of the magnetic gluons, 
the gap from QCD in weak coupling limit could be large as compared 
to that obtained under the condition which the magnetic gluons 
acquire static screening mass. 
If we extrapolate the gap down to the intermediate region of the 
chemical potential, it increases up to 100 MeV. 
However, as temperature grows, the color superconducting gap 
is reduced, then, the phase undergoes phase transition. 
The order of the transition seems to be second order in 
either case of with or without gauge fields. 
However, with gauge fields, there is a possibility of 
first order or a smooth crossover.

We consider the color superconductivity under the situation where 
a flavor asymmetry takes place.  The color superconductivity 
at finite flavor asymmetry has been studied in 
Ref. \cite{alford,bedaque}. Especially, in Ref. \cite{bedaque}, 
it has been argued that a mixed phase of 
the normal and the superconducting phase exists 
with a sufficiently large flavor asymmetry. 
In this paper, we study at finite temperature and discuss 
what extent does the mixed phase persist. 
Such an investigation has importance in studies of neutron stars and, 
of course, high-energy heavy-ion collisions at 
Alternating Gradient Synchrotron (AGS) 
and CERN Super Proton Synchrotron (SPS).

The study of the color superconductivity in QCD is nontrivial 
and, therefore, we choose a tractable model to describe 
the quark--quark condensate. 
The most familiar of such models is known as 
the NJL (-type) model. Particularly, in the leading order of the 
$1/N_c$ expansion, the resultant gap function is 
independent of momentum and it makes our analysis simple one. 
Moreover, the structure of the color superconducting 
phase depends on the number of the quark flavors. 
At extremely high density, where we can neglect the mass of $s$ quark, 
the color-flavor locked phase appears \cite{cfl}. However we expect, 
at relatively low density, there is a window of two--flavor 
color superconducting phase. 
Such being the case, we use the two--flavor, three--color NJL-type model.

This paper is organized as follows. 
In Sec. II, we present our model and derive 
the thermodynamic potential at finite density, temperature and 
flavor asymmetry. 
Then, in Sec. III, we present numerical results for 
the thermodynamic potential. Section IV is devoted to the 
summary and discussion. Our unit are $\hbar=k_B=c=1$.

\section{MODEL}

Our aim is to explore the phase diagram of two--flavor QCD. 
The $SU(2)_L \times SU(2)_R$ chially symmetric Lagrangian \cite{njl}
\begin{eqnarray}
{\cal L}_{NJL}&=&\bar{\psi}i\gamma\cdot\partial\psi+{\cal L}_{int},\\
{\cal L}_{int}&=&G_1\left[(\sigmac)^2+(\pic)^2\right],\label{eqn:lag}
\end{eqnarray}
is our starting point. The quark field $\psi$ is assigned to the 
fundamental representation of the color $SU(N_c)$ group and 
carries the flavor index $i$ and 
the color index $a$ as $\psi_{a}^{i}$. 
The Pauli matrices $\vec{\tau}=(\tau_1,\tau_2,\tau_3)$ act 
in the flavor space. 
In Eq. (\ref{eqn:lag}), we include two four--fermion interactions, 
namely, the scalar isosinglet $(\sigma)$ and 
pseudo-scalar isotriplet $(\pi)$. 
However, we do not abandon the possibility of 
adding other four--fermion interactions; e.g., the scalar isotriplet 
$(a_0)$ and the pseudo-scalar isosinglet $(\eta')$. 

Applying the Fierz transformations to 
these color singlet interactions, one can transform them 
into color $\bvec{\bar{3}}$ and $\bvec{6}$ channels. 
In this paper, we consider the following 
scalar isoscalar color $\bvec{\bar{3}}$ 
diquark condensation which breaks $SU(3)_C$ to $SU(2)_C$; i.e., 
taking a third direction as a preferred one in color space, 
\begin{eqnarray}
\Delta=\Delta_3=\Delta_3^{\ast} 
\sim \epsilon^{ij}\epsilon_{ab3}\langle(\psi_a^i)^{T}C\gamma_5\psi_b^j\rangle,
\end{eqnarray}
where $C$ is a charge conjugation matrix, defined by 
$C\gamma_{\mu}C^{-1}=-\gamma_{\mu}^T$ and $C=-C^{-1}=-C^T$ 
and $\Delta$ is independent of momentum in our approximation 
(mean field approximation or, equivalently, 
leading order in the $1/N_c$ expansion). 
The interaction Lagrangian which leads to 
the diquark condensation is written as
\begin{eqnarray}
{\cal L}_{diq}=G_2\left[-(\bar{\psi}\gamma_5\tau^2\lambda^2\psi^C)
          (\bar{\psi}^C\gamma_5\tau^2\lambda^2\psi)+\cdots\right],
\label{eqn:lag2}
\end{eqnarray}
where $\tau^2$ is the antisymmetric Pauli matrix, $\lambda^2$ is 
the antisymmetric (color  $\bvec{\bar{3}}$) color generator, and 
the ellipsis denotes irrelevant diquark channels. 
Since the NJL model is not renormalizable, one cannot define the 
model until a regularization scheme has been specified. 
In this paper, we calculate the thermodynamic potential with a 3D 
sharp cutoff in momentum space. 
For a mesonic sector, we use a set of parameters: 
a coupling constant $G_1=5.01~{\rm GeV}^{-2}$ and 
a 3D ultraviolet cutoff $\Lambda=0.65$ GeV \cite{klevansky}. 
These parameters reproduce the value of $f_{\pi}=93$ MeV and 
the quark condensate $\langle\bar{q}q\rangle=(-225{\rm MeV})^{-3}$. 
The coupling constant $G_2$, mediating an 
attractive four--fermion diquark interaction, is related to $G_1$ 
through the Fierz transformations. 
In other words, the strength of $G_2$ is fixed by 
the Fierz transformations. The Lagrangian (\ref{eqn:lag}) gives 
$G_2=G_1/4$. However, as mentioned before, one can change 
the ratio $G_2/G_1$ by adding other color singlet channel to 
Eq. (\ref{eqn:lag}). Therefore we take $G_2$ independent on $G_1$ and 
study the $G_2$ dependence of $\Delta$ and a phase structure.

We start our study with an investigation of the thermodynamic potential 
as a function of $\Delta$ at nonzero temperature, 
``averaged'' chemical potential and ``isospin'' chemical potential. 
The averaged chemical potential $\mu_B$ and isospin 
chemical potential $\mu_I$ are given as follows
\begin{eqnarray}
\mu_B=\frac{\mu_u+\mu_d}{2}~~,~~\mu_I=\frac{\mu_u-\mu_d}{2},
\end{eqnarray}
where $\mu_{u(d)}$ is a chemical potential of the $u(d)$ quark. 
In order to evaluate the thermodynamic potential, 
one can use auxiliary field method with the saddle point 
approximation or the Cornwall-Jackiw-Tomboulis effective action 
\cite{cjt} up to two-loop level with the Nambu-Gorkov formalism. 
These two methods are correspond to 
the evaluation at the leading order of the $1/N_c$ expansion 
and consistent with each other. 
The thermodynamic potential $V(\Delta;\mu_B,\mu_I)$ 
in the leading order of the $1/N_c$ expansion is obtained as follows
\begin{eqnarray}
V(\Delta;\mu_B,\mu_I)&=&\frac{\Delta^2}{4G_2}
+i\int\frac{d^4q}{(2\pi)^4}\bigg{\{}
\ln\left[\left((\omega+\mu_I)^2-\bvec{q}^2
-\mu_B^2\right)^2-4\mu_B^2\bvec{q}^2\right]\nonumber\\
&&+\ln\left[\left((\omega-\mu_I)^2-\bvec{q}^2
-\mu_B^2\right)^2-4\mu_B^2\bvec{q}^2\right]\nonumber\\
&&+2\ln\left[\left((\omega+\mu_I)^2-\bvec{q}^2
-\mu_B^2-\Delta^2\right)^2-4\mu_B^2\bvec{q}^2\right]\nonumber\\
&&+2\ln\left[\left((\omega-\mu_I)^2-\bvec{q}^2
-\mu_B^2-\Delta^2\right)^2-4\mu_B^2\bvec{q}^2\right]\bigg{\}}.
\end{eqnarray}
The finite temperature thermodynamic potential 
$V(\Delta;T,\mu_B,\mu_I)$ is obtained by using the formula:
\begin{equation}
\int\frac{d^4q}{(2\pi)^4}~F(q) \to \frac{-1}{4\pi}
\int\frac{d^3q}{(2\pi)^3}\oint d\omega~
\tanh\frac{\beta\omega}{2}F(\omega,\bvec{q}),
\end{equation}
where $\beta=1/T$ and the $\omega$-integration is to be 
understood that it is round an anti-clockwise contour including 
the pole of $F(\omega,\bvec{q})$. Thus, we obtain 
the thermodynamic potential as follows
\begin{eqnarray}
V(\Delta;T,\mu_B,\mu_I)&=&\frac{\Delta^2}{4G_2}
-2\int^{\Lambda}\frac{q^2dq}{2\pi^2}\bigg{\{}
\epsilon_{-}(q)+\epsilon_{+}(q)\nonumber\\
&+&T\ln\left(1+\exp\left[-\beta(\epsilon_{-}(q)-\mu_I)\right]\right)
+T\ln\left(1+\exp\left[-\beta(\epsilon_{-}(q)+\mu_I)\right]\right)\nonumber\\
&+&T\ln\left(1+\exp\left[-\beta(\epsilon_{+}(q)-\mu_I)\right]\right)
+T\ln\left(1+\exp\left[-\beta(\epsilon_{+}(q)+\mu_I)\right]\right)
\bigg{\}}\nonumber\\
&-&4\int^{\Lambda}\frac{q^2dq}{2\pi^2}\bigg{\{}E_{-}(q)+E_{+}(q)\nonumber\\
&+&T\ln\left(1+\exp\left[-\beta(E_{-}(q)-\mu_I)\right]\right)
+T\ln\left(1+\exp\left[-\beta(E_{-}(q)+\mu_I)\right]\right)\nonumber\\
&+&T\ln\left(1+\exp\left[-\beta(E_{+}(q)-\mu_I)\right]\right)
+T\ln\left(1+\exp\left[-\beta(E_{+}(q)+\mu_I)\right]\right)\bigg{\}},
\end{eqnarray}
where
\begin{eqnarray}
\epsilon_{\pm}(q)=|\bvec{q}|\pm\mu_B
\end{eqnarray}
are the gapless (color 3) 
quasi-particle energies relative to the (averaged) Fermi surface and
\begin{eqnarray}
E_{\pm}(q)=(|\bvec{q}|\pm\mu_B)
\sqrt{1+\frac{\Delta^2}{(|\bvec{q}|\pm\mu_B)^2}},
\end{eqnarray}
are the gapful (color 1, 2) quasi-particle energies 
and 3D cutoff $\Lambda$ is introduced to regularize the divergent integrals. 

The extremum condition for $V(\Delta;T,\mu_B,\mu_I)$ with respect to 
$\Delta$ lead to the Schwinger-Dyson equation (SDE) for $\Delta$ at nonzero 
temperature, density, and flavor asymmetry. 
The Cooper pairs condensate $\langle\psi\psi\rangle$ is 
related $\Delta_0$ the solution of the SDE as
\begin{eqnarray}
\langle\psi\psi\rangle=\Delta_0/(2G_2).
\end{eqnarray}
The ``baryon number density'' and the ``isospin density'' 
are obtained by
\begin{eqnarray}
n_B & \equiv & n_u + n_d = 
-\frac{\partial}{\partial\mu_B}V(\Delta;T,\mu_B,\mu_I)\nonumber\\
n_I & \equiv & n_u - n_d = -\frac{\partial}{\partial\mu_I}
V(\Delta;T,\mu_B,\mu_I).
\end{eqnarray}

\section{COLOR SUPERCONDUCTIVITY WITH ASYMMETRIC MATTER}

Figure 1 shows the $G_2$ and $\mu_B$ dependence of $\Delta$ 
at zero temperature and zero flavor asymmetry. 
The value $G_2=3G_1/4$ was taken by Berges and Rajagopal
\cite{rajagopal}, 
where $G_1$ is the coupling constant in $\sigma$ (scalar isosinglet) channel. 
Note that we use the $G_1$ different from 
Berges and Rajagopal and 3D sharp cutoff instead of 
introducing a smooth form factor $F(q)=\Lambda^2/(q^2+\Lambda^2)$. 
Then, the behavior of $\Delta$ slightly different from theirs. 
The value $G_2=G_1/2$ and $G_2=G_1/4$ are introduced by 
the instanton vertex of two--flavor QCD and 
the $SU(2)_L \times SU(2)_R$ symmetric NJL model, respectively. 
As we can see, $\Delta$ is quite sensitive to $G_2$. 
For $G_2=3G_1/4$, $\Delta$ could be as large as 100 MeV. 
However, it is about 50 MeV for $G_2=G_1/2$ and 
negligible small for $G_2=G_1/4$. 
In the following, we use $G_2=3G_1/4$ to get sizeable gap.

Figure 2 shows the zero temperature and $\mu_B=400$ MeV thermodynamic 
potential as a function of $\Delta$ for several $\mu_I$. 
For $\Delta$ larger than $\mu_I$, all curves are coincident 
with each other. 
This is because, for $\Delta > \mu_I$, 
the flavor asymmetry does not affect the thermodynamic potential. 
As $\mu_I$ grows, the potential in the trivial state $(\Delta=0)$ 
decreases. It has two degenerate minima and the superconducting phase 
and the normal phase coexist at some value of $\mu_I$. 
For more large values of $\mu_I$, the normal phase is favored. 
As a result, $\Delta$ behaves constant up 
to a certain critical value of $\mu_I$ and, then, jumps to zero. 
We have a strong first order phase transition. 
We can see the same behavior of the thermodynamic potential 
and the order parameter in chiral symmetry restoration 
at zero temperature and finite baryon density.

In Fig. 3, we present the phase diagram as a function of $\mu_B$ and 
$\mu_I$ at zero temperature. Along the horizontal axis, 
color superconducting phase exists for arbitrary nonzero $\mu_B$ 
because the BCS instability exists for arbitrary small coupling 
strength. On the other hand, as mentioned before, 
we know that it undergoes a phase transition of first order with 
increasing flavor asymmetry. Then we get the phase boundary 
such as shown in Fig. 3. This boundary reflect the $\mu_B$ dependence 
of $\Delta$ at $\mu_I=0$ (see Fig. 1). 
It is also interesting to see the phase diagram as a function of 
$\mu_B$ and $\mu_I$. 
Figure 4 shows the phase diagram in the $n_B$-$n_I$ plane. 
For the flavor asymmetry $n_I/n_B$ larger than about 15\% the system 
remains in the mixed phase for arbitrarily high densities. 
This nature is almost the same as shown in Ref. \cite{bedaque}. 
For the flavor asymmetry larger than about 40\% 
the phase transition is complete and the system undergoes a 
phase transition into the normal phase. 
This point is differ from the results of Ref. \cite{bedaque} 
and the difference arises from the choice of the parameters.

We now turn to physics at finite temperature. 
Figure 5 shows the $\mu_I$ dependence of $V$ for $T=20$ MeV 
and $\mu_B=400$ MeV. 
At relatively low temperature, a first order 
phase transition occurs. On the other hand, 
at relatively high temperature, the phase transition changes 
to into second order. From the behavior of the thermodynamic 
potential, the existence of a tricritical point in the 
$T$-$\mu_I$ plane is expected. 
In Fig. 7, we present the typical phase diagram as a function of $T$ and 
$\mu_I$ for $\mu_B$ fixed to 400$/$500 MeV. 
It is known that, in four--fermion interaction model which neglects the 
gauge fields, we have a second order phase transition for $\mu_B \neq 0$ 
and $\mu_I=0$ with increasing temperature 
(BCS theory predicts the critical temperature $T_c$ is given 
by $T_c \sim 0.57\Delta_{T=0}$). 
On the other hand, we have a first order transition 
for $T=0$ with increasing flavor asymmetry $\mu_I$. 
Therefore the existence of a tricritical point 
in the $T$-$\mu_I$ plane , 
where the second order phase transition becomes that of the first order, 
is expected. 
In fact, we found 
\begin{eqnarray}
T& \sim & 35~~{\rm MeV},~~\mu_I \sim 65~~{\rm MeV}~~{\rm for}~\mu_B=400~~{\rm 
MeV},\nonumber\\
T&\sim & 42~~{\rm MeV},~~\mu_I \sim 69~~{\rm MeV}~~{\rm for}~\mu_B=500~~{\rm
MeV},\nonumber
\end{eqnarray}
from the condition
\begin{eqnarray}
a_2(T,\mu_I)=a_4(T,\mu_I)=0,
\end{eqnarray}
where we expand the thermodynamic potential, in the vicinity of the 
tricritical point, as a power series in $\Delta$:
\begin{eqnarray}
V(\Delta;T,\mu_I)=V(0,T,\mu_I)+a_2(T,\mu_I)\Delta^2
                  +a_4(T,\mu_I)\Delta^4+\cdots.
\end{eqnarray}
Therefore, there is a possibility of a first order 
phase transition as temperature grows, with 
sufficiently large flavor asymmetry.

\section{SUMMARY AND DISCUSSION}

In summary, we studied the phase structure of 
the color superconductivity at finite density, temperature and 
flavor asymmetry by investigating the thermodynamic 
potential in the NJL model. 

Since specifying the ratio $G_2/G_1$ by the Fierz 
transformations in the mean-field approximation 
is not unique, we examined the $G_2$ dependence 
of $\Delta$ and found that $\Delta$ is quite sensitive to $G_2$.

First, let us see $\mu_I$ dependence of $\Delta$. 
At zero temperature, as $\mu_I$ grows, 
the superconducting gap $\Delta$ remains constant 
until $\mu_I$ is increased to its critical value and, then, vanishes. 
This behavior seems to be the same as obtained by Bedaque \cite{bedaque}. 
Following Alford {\it et al.}, let us consider a concrete example; 
a gold nucleus at a baryon density eight times in nuclear matter. 
This matter, the number of $d$ quark exceed that of $u$ quark, 
corresponds to $\mu_B=535$ MeV and $\mu_I=12$ MeV (up to sign). 
They have found that the gap is a little (about 18\%) decreased. 
Our resultant gap does not changed in this condition. 
The distinction maybe arise from the difference in 
the coupling constant and the regularization scheme. 

Second, concerning the phase structure, 
we found the similar phase structure in the $n_B$-$n_I/n_B$ plane to 
that reported by Bedaque \cite{bedaque}. 
Moreover, we found that the structure of the color superconducting phase 
is remarkably rich:\\
(1) mixed phase appears at sufficiently large flavor asymmetry 
with temperature less than 40 MeV,\\
(2) tricritical point exists in the $T$-$\mu_B$-$\mu_I$ phase diagram.\\
The most favorable condition for the color superconductivity, 
cold and dense quark matter, are at the core of neutron stars. 
However, the values of $T$, $\mu_B$ and $\mu_I$ accomplished in 
heavy-ion collisions at, for example, 
AGS and SPS, maybe reach into the mixed phase, 
since it lies up until $T\simeq 40$ MeV (see Fig. 7). 
Then, if the charged bubbles is formed, it may lead 
to observable effects, e.g., an obvious 
change of the $\pi^{-}/\pi^{+}$ ratio.

Our results strongly depend on the regularization scheme, although 
qualitative features do not changes. 
In addition, it seems to be necessary to take into account 
next-to-leading order in the $1/N_c$ expansion, 
because the ratio of $G_2/G_1$ has a order ${\cal O}(1/N_c)$. 
Also, the dynamical chiral symmetry breaking and 
the pion condensation should be included in further study. 
The pion condensed phase has been confirmed by 
the Monte Carlo simulation of two--color QCD \cite{fid}. 
We also plan to study 
the effective potential within QCD in weak coupling limit and 
the three flavor case where the color flavor locked phase takes place.

\newpage

\begin{figure}
\begin{center}
\leavevmode
\epsfxsize=10cm
\epsfbox{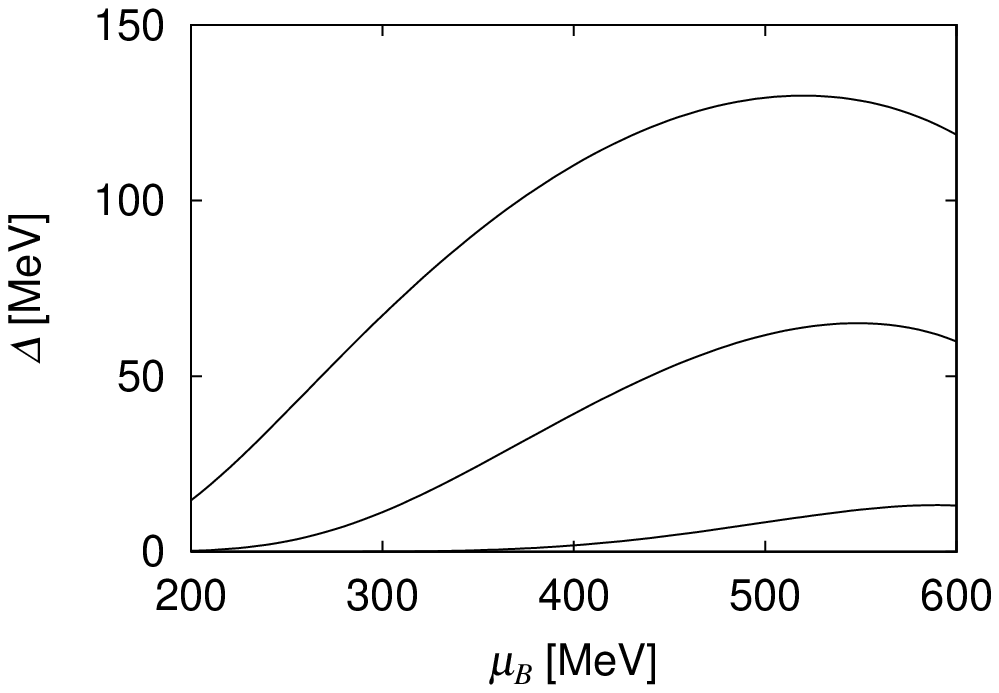}
\end{center}
\leavevmode
\caption{Superconducting gap $\Delta$ as a function of $\mu_B$. The 
curves correspond to $G_2/G_1=3/4,1/2,1/4$ (top to bottom).}
\end{figure}

\begin{figure}
\begin{center}
\leavevmode
\epsfxsize=10cm
\epsfbox{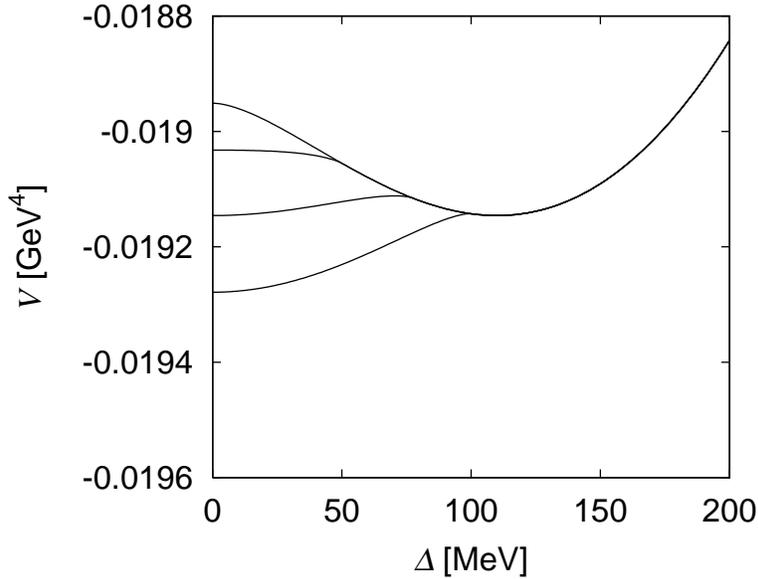}
\end{center}
\leavevmode
\caption{The zero temperature thermodynamic potential as a function of
 $\Delta$ for several $mu_I$. The curves correspond to 
$\mu_I=0,50,77.3,100$ MeV. The value $\mu_I=77.3$ MeV corresponds to 
a critical isospin chemical potential. 
Note that we neglect the contribution of color 3 quark to $V$ 
which is independent of $\Delta$. The same neglect will be done in 
Fig. 5 and 6.}
\end{figure}

\begin{figure}
\begin{center}
\leavevmode
\epsfxsize=10cm
\epsfbox{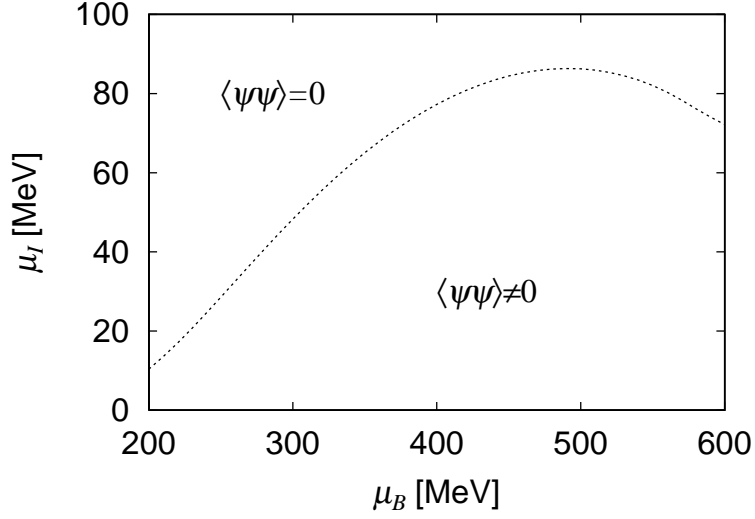}
\end{center}
\leavevmode
\caption{Phase diagram in the $\mu_B$-$\mu_I$ plane. The (dotted) line
 indicates the first order phase transition.}
\end{figure}

\begin{figure}
\begin{center}
\leavevmode
\epsfxsize=10cm
\epsfbox{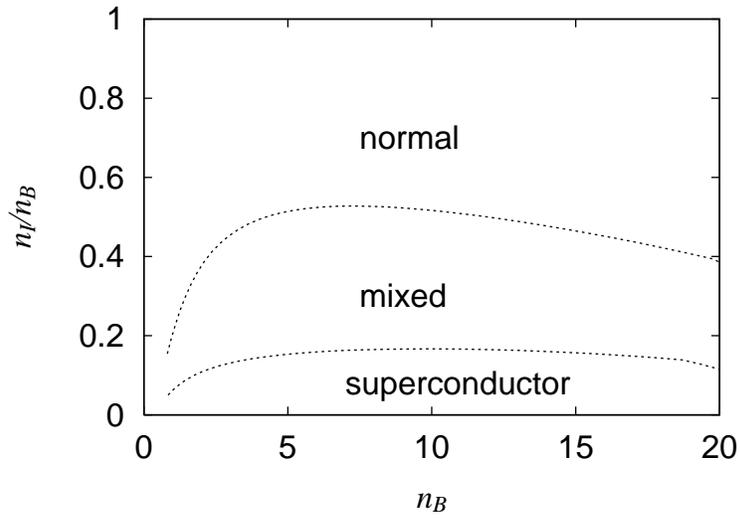}
\end{center}
\leavevmode
\caption{Phase diagram in the $n_B$-$n_I/n_B$ plane 
(in unit of $10^6{\rm MeV}^3$).}
\end{figure}

\begin{figure}
\begin{center}
\leavevmode
\epsfxsize=10cm
\epsfbox{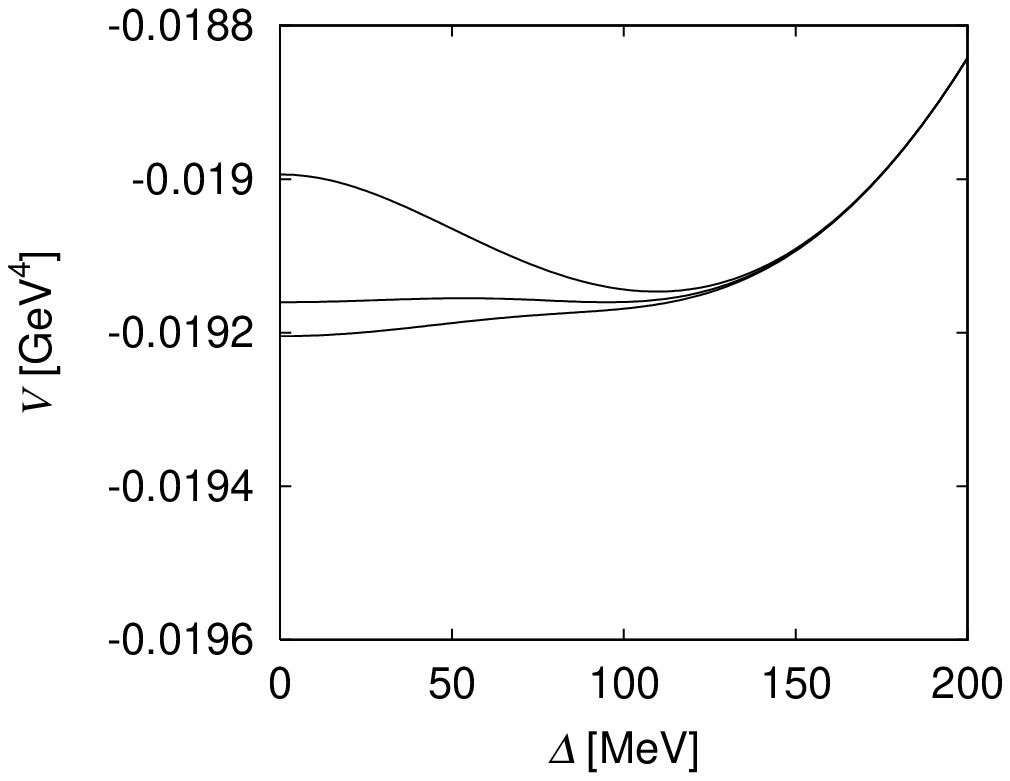}
\end{center}
\leavevmode
\caption{The thermodynamic potential for $T=20$ MeV, $\mu_B=400$ MeV and 
several $\mu_I$. The curves correspond to $\mu_I=0,71,80$ MeV 
(top to bottom).}
\end{figure}

\begin{figure}
\begin{center}
\leavevmode
\epsfxsize=10cm
\epsfbox{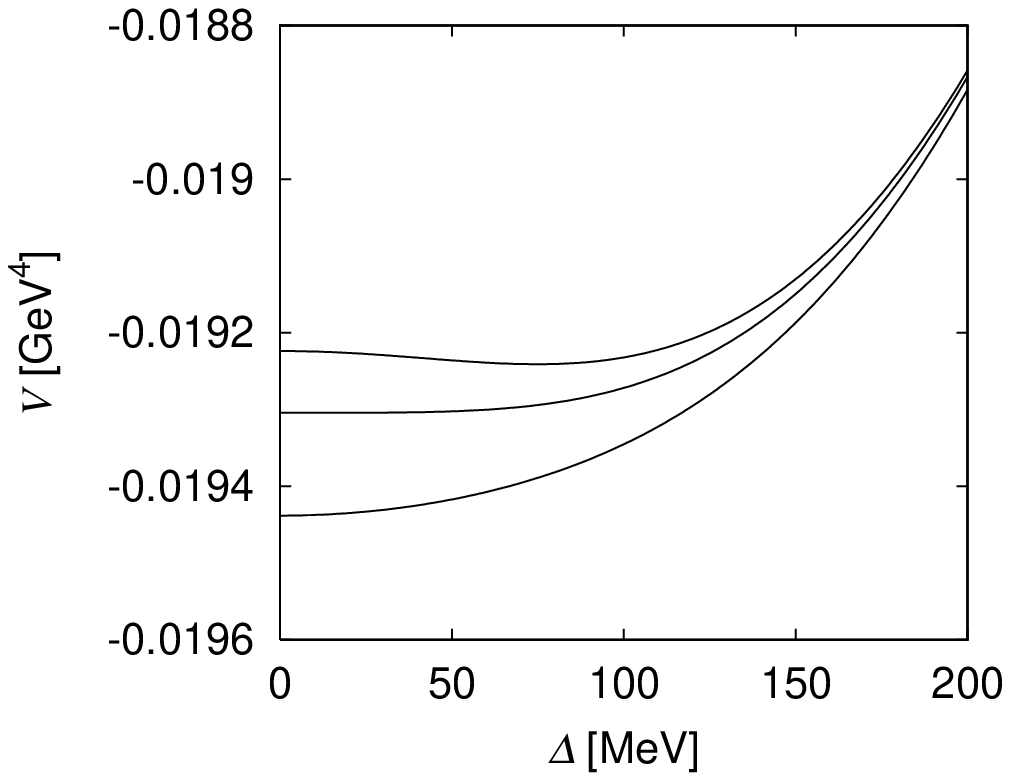}
\end{center}
\leavevmode
\caption{The thermodynamic potential for $T=50$ MeV, $\mu_B=400$ MeV and 
several $\mu_I$. The curves correspond to $\mu_I=0,49,80$ MeV 
(top to bottom).} 
\end{figure}

\begin{figure}
\begin{center}
\leavevmode
\epsfxsize=10cm
\epsfbox{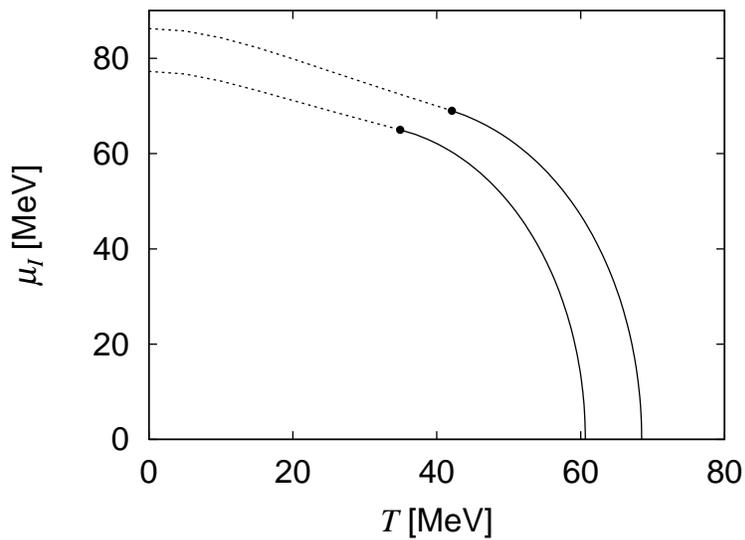}
\end{center}
\leavevmode
\caption{Phase boundaries in the $T$-$\mu_I$ plane for 
$\mu_B=500$ MeV (upper line) and $400$ MeV (lower line). 
The solid lines describe the 
transition of second order and the dashed line describe that of 
first order. Points indicate tricritical points.}
\end{figure}

\end{document}